\documentclass{article}

\usepackage{amssymb,amsmath}
\usepackage{times}

\usepackage{graphicx}

\usepackage{amssymb}

\begin{document}
\centerline{\Large{Attosecond neutron Compton scattering from
   protons}}
\centerline{\Large{entangled with adjacent electrons}}

\vspace{1cm}
\centerline{C. Aris Chatzidimitriou-Dreismann$^1$}

\vspace{1cm} \noindent $^1$ Institute of Chemistry, Sekr C2,
Technical University Berlin, D-10623 Berlin, \\Germany.
\textit{Email:} dreismann@chem.tu-berlin.de
%

\vspace{1cm} \noindent \abstract{The effect of "anomalous"
scattering of neutrons and electrons from protons in the
electron-volt energy-transfer range is considered, and related
experimental results are mentioned. A recent independent
confirmation of this effect with a new data analysis procedure is
presented. Due to the very short characteristic scattering time,
there is no well defined
 separation of time scales of electronic and protonic motions.
An outline of a proposed theoretical interpretation is presented,
which is based on the fact that scattering protons  represent
\textit{open} quantum systems, thus being subject to decoherence.
}\\





\section{Introduction}\label{intro}

Several neutron Compton scattering (NCS) experiments on liquid and
solid samples containing protons or deuterons show a striking
anomaly, which is a shortfall in the intensity of epithermal
neutrons scattered by the protons and deuterons. E.g., neutrons
colliding with water for just 100-500 attoseconds (1 as =
$10^{-18}$\,s) will see a ratio of hydrogen to oxygen of roughly
1.5 to 1, instead of  2 to 1 corresponding to the chemical formula
H$_2$O; cf. \cite{PRL97,highlights}.  The experiments were done at
the ISIS neutron spallation facility, Rutherford Appleton
Laboratory, UK. Due to the large energy and momentum transfers
applied, the duration of a neutron-proton scattering event is a
fraction of a femtosecond which is extremely short compared to
condensed-matter relaxation times.

This new effect has been confirmed using an independent method,
electron-proton Compton scattering (ECS), at the Australian
National University. \cite{PRL03,NIMB}. ECS experiments from a
solid polymer showed the exact same shortfall in scattered
electrons (with energy about 20-35 keV; scattering angle
45$^\circ$) from hydrogen nuclei, comparable to the shortfall of
scattered neutrons in accompanying NCS experiments on the same
polymer. The similarity of the results is striking because the two
projectiles interact with protons via fundamentally different
forces -- electromagnetic and strong \cite{PRL03,NIMB,highlights}.


Due to its novelty and far-reaching consequences, however, this
effect has been the focus of various criticisms,
cf.~\cite{Blostein,Cowley}. Therefore,
 considerable work to identify possible sources of experi\-men\-tal
and data-analysis errors was made during the last five years,
which succeeded to demonstrate the excellent working conditions of
the spectrometer Vesuvio at ISIS \cite{J+T}. Moreover, extending
these investigations,  the complete "exact formalism" of data
analysis \cite{Blostein} was applied to NCS-data by Senesi et al.
\cite{Senesi}, for the first time;
 analysis of time-of-flight (TOF) spectra from solid HCl
 revealed the existence of a strong "anomalous" decrease
of the scattering intensity from H (up to  34\%). Additionally,
this result was found to be in excellent agreement with the
corresponding outcome of  the standard data analysis procedure
applied at ISIS \cite{Senesi}.

Recently, scattering of neutrons in the 24-150 keV incident energy
range from H$_2$O relative to that of D$_2$O was investigated
\cite{Moreh2005}. In clear contrast to the NCS and ECS results, it
was claimed that the measured neutron scattering intensity ratios
exhibit no anomalous behaviour. However, an improved analysis
\cite{antiMoreh} of the keV-data within the frame of standard
theory showed that the considered scattering anomaly is present at
both 5-100 eV  and the keV ranges of neutron energies.

Very recently, the mentioned standard NCS-data analysis method
\cite{J+T} was successfully compared with a newly proposed (by B.
Dorner, ILL) model-free data-analysis procedure, the latter being
independent of the form of the momentum distribution and the
resolution function \cite{PRBDorner}. In this work, the original
results from the metallic hydride NbH$_{0.8}$ \cite{EPL99} were
analyzed.  The comparison of results obtained with the mentioned
two independent methods underline the importance of the effect
under consideration; see Fig.~1.

\begin{figure}[htb]
  \centering
  \includegraphics[width=70mm]{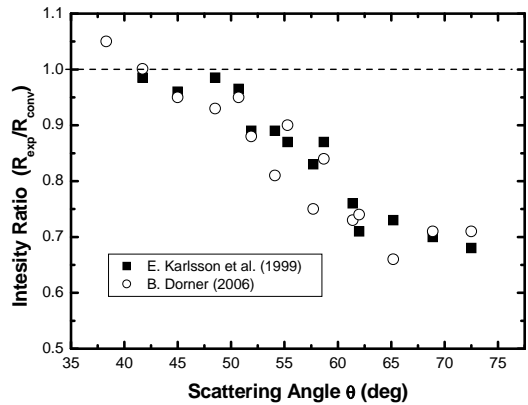}
\caption[]{"Anomalous" scattering from the solid metallic hydride
NbH$_{0.8}$ \cite{EPL99}. Shown is the measured ratio $R_{exp}=
\sigma_H/\sigma_{Nb}$ of scattering cross-sections of H and Nb
normalized with their expected (tabulated) ratio $R_{conv}$.
Broken line: conventional expectation; "full" symbols: results
taken from the original publications \cite{EPL99}; "open" symbols:
results of the model-independent data analysis procedure invented
by Dorner \cite{PRBDorner}(a).  } \label{Fig2}
\end{figure}

\section{On scattering time}

In the context of  NCS, as provided by the Vesuvio setup, the
Impulse Approximation (IA) is valid \cite{Sears,Watson} and the
characteristic time scale --- often termed "scattering time",
$\tau_{sc}$ --- of the neutron-proton scattering process is very
short \cite{PRL03,J+T,EPL99,JACS},
\begin{equation}
 \tau_{sc} \sim 100 - 1000 \ {\rm as}\
\label{subfemto}
\end{equation}
(as: attosecond). This is a consequence of the large energy (up to
100 eV) and momentum transfers attained with the Vesuvio
instrument \cite{J+T} and
 follows from the theoretical result valid in the IA \cite{Sears,Watson}
\begin{equation}
 \tau_{sc}\, q\, v_0 \approx 1 \ ,
\label{tau}
\end{equation}
 where $v_0$
is the root-mean-square (rms) velocity of the nucleus and
$\hbar q$ is the (absolute value of) momentum transfer from 
 the neutron to the proton.
The time $\tau_{sc}$ is given by the $t$-width of the intermediate
correlation function $F(q,t)$, which is related to the dynamic
structure factor $S(q,\omega) $ by Fourier transform
\begin{equation}
S(q,\omega) = \frac{1}{2\pi}\int^\infty _{-\infty} \exp{(-i\omega
t )}F( q,t)\,dt \ .
  \label{F}
\end{equation}

It is interesting to note that the "actual duration" $\tau_{act}$
of a neutron-proton interaction should be even shorter, as a
classical estimate indicates. E.g. a neutron with kinetic energy
$E_0 \approx 10$ eV will pass a distance of $10^{-5}$ \AA \, (i.e.
the range of the strong interaction) in a much shorter time,
$\tau_{act} \approx 10^{-19}$ s. In the light of Relativity
Theory, this has a crucial consequence: the "actual" (or
"effective") scattering system  --- i.e. a proton and its adjacent
electrons --- has a linear dimension not larger than
$$
\Delta s_{max} =c \cdot \tau_{act} \lesssim 0.3\, {\rm \AA  }
 $$
($c$: velocity of light), since the neutron-nucleus scattering
dynamics during $\tau_{act}$ cannot be  causally influenced by
other particles being more than $\Delta s_{max}$ apart from the
nucleus. Consequently, the scattering system must necessarily be
of "microscopic" dimensions; it contains the scattering nucleus
and a part of the adjacent electron density.

However, this is not in conflict with the above estimate, for the
following reason. As standard theory shows \cite{Watson},
$S(q,\omega)$ is peaked around the nuclear recoil energy $E_q=
\hbar^2q^2/2m$. The scattering time $\tau_{sc}$ is also given by
the inverse of the width $\Delta E$ of $S(q,\omega)$, and
$S(q,\omega)$ plays the role of the probability density
distribution for transferring energy $\hbar\omega$ from the
neutron to the proton, when the momentum transfer is $\hbar q$.
That is,  $\tau_{sc} \approx \hbar / \Delta E$.
 (Interestingly, as Gidopoulos  \cite{Nikitas} showed,
  $\tau_{sc}$ is also
  about the inverse of the energy spread of the proton wavefunction
  after collision.)
  For a typical value $\Delta
E\approx 10$ eV, one gets $\tau_{sc} \approx 10^{-16} - 10^{-17}$
s.
  In other words,  the scattering time $\tau_{sc}$
 gives {\it a statistical measure of the length of the time interval
during which an actual neutron-proton collision may occur} --- in
the same way that the spatial extent of a particle wavefunction
 gives a statistical measure of the extent of the region in which
the particle may be found.

To shed more light upon the issue of "relevant scattering time",
one may also refer to the celebrated Margolus-Levitin theorem
\cite{Margolus}. Let us consider the neutron-proton system during
the collisional process. Obviously, the initial and final states
of it are  very different and so they can  safely be assumed to be
orthogonal. This theorem asserts that in takes at least a time
$T_\perp \geq (\pi \hbar) / (2 E_s) $ for the system to evolve
from its initial to any orthogonal final state.   $E_s$ is the
system's average energy minus its ground state energy. $T_\perp$
provides a strict bound for the considered dynamical process
\cite{Margolus}.
 Note that in NCS one has $E_s \approx \hbar\omega$ with $\hbar\omega$ taken at the
 peak center, and thus
$E_s$ is of similar order as the aforementioned energy spread
$\Delta E$. Thus it is revealing that also this time $T_\perp$ is
very similar to the aforementioned scattering time, i.e. $ T_\perp
\lesssim \tau_{sc}$.

\section{Theoretical model --- NCS from open quantum systems}

In the following we present an outline of a recently proposed
theoretical interpretation of the considered effect, which is
based on the general theory of scattering from open quantum
systems \cite{Aris+Stig}. As explained above, the scattering
system must necessarily be of "microscopic" dimensions (i.e. it is
of the order of one \AA \,  or less), and since it is embedded in
condensed matter it represents an "open" quantum system
\cite{Petruccione}. This point is crucial, since standard neutron
scattering theory always assumes a condensed matter scattering
system to be closed; see e.g. \cite{vHove,Squires}.

\subsection{Scattering from closed systems}

First let us consider neutron scattering from a {\it closed}
system consisting of $N$ particles with the same scattering length
$b$, and the
 N-body Hamiltonian  $H_{total}=H_{0}+V$ with the
interaction
\begin{equation}
V(\mathbf{r})=\lambda \,n(\mathbf{r}) ,\ \ \ \ \ {\rm with} \ \ \
\ \
 \lambda =\frac{2\pi \hbar ^{2}}{m}\,b \  .
     \label{a1}
\end{equation}
$m$ is the neutron mass, $n(\mathbf{r})$ is the particle density
operator
\begin{equation}
n(\mathbf{r})=\frac{1}{V}\sum_{j=1}^{N}\delta
(\mathbf{r}-\mathbf{R}_{j}) \ ,
   \label{a1-1}
\end{equation}
where $V$ is the volume, and  $\mathbf{R}_{j}$ is the  position of
the $j$-th particle; cf. the textbook \cite{Squires}.

In the \textit{interaction} picture, the Schr\"{o}dinger equation
is now (setting for simplicity $\hbar =1$) $ i\partial _{t}\Psi
=\lambda \,n(\mathbf{r,}t)\Psi$,  with the perturbative solution
\begin{equation}
\Psi (t)=\Psi (0)-i\lambda \int_{0}^{t}n(\mathbf{r},t^{\prime })\,
dt^{\prime }\Psi (0).  \label{a2}
\end{equation}

We write the transition probability $W(t)$ between initial states
$\psi _{i}$ (with probability $P_{i}$) and final states $\psi
_{f}$ of the scattering system to be given by
\begin{equation}
W(t)=\sum_{i,f}\mid \langle \psi _{f}\mid \lambda
\int_{0}^{t}n(\mathbf{r}, t^{\prime })\, dt^{\prime }\mid \psi
_{i}\rangle \mid ^{2}P_{i}.  \label{a3}
\end{equation}
It should be noted that $\psi _{i}$ and $\psi _{f}$ are
eigenstates of the $N $-body Hamiltonian $H_{0}$ omitting the
probe system \cite{vHove,Squires}. The transition probability is
then given in the form
\begin{equation} W(t)=\lambda
^{2}\int_{0}^{t}dt^{\prime }\int_{0}^{t}dt^{\prime \prime
}\sum_{f}\langle \psi _{f}\mid n(\mathbf{r,}t^{\prime })\,\rho \,n(\mathbf{r,%
}t^{\prime \prime })\mid \psi _{f}\rangle ,  \label{a3+1}
\end{equation}
with
 $ \rho =\sum_{i}\mid \psi _{i}\rangle P_{i}\langle \psi_{i}
 \mid$,
where we have noted that $n^{\dag}(\mathbf{r},t)=n(\mathbf{r},t)$.

In an actual scattering experiment from condensed matter, we do
not measure the cross-section for a process in which the
scattering system goes from a specific initial state $\psi _{i}$
to another state $\psi _{f}$, both being unobserved states of the
many-body system. Therefore, one takes an appropriate average over
all these states \cite{vHove,Squires}, as done in Eq.~(\ref{a3}).

Given the initial ($\mathbf{k}_{0}$) and final ($\mathbf{k}_{1}$)
momenta of an impinging neutron and introducing the momentum
transfer $\mathbf{q}= \mathbf{k}_{0}-\mathbf{k}_{1}$ from the
probe particle to the scattering system, the Fourier transform of
the particle density reads
\begin{equation}
n(\mathbf{r},t)=\frac{1}{(2\pi )^{3}}\int
\!d\mathbf{q}\,n(\mathbf{q},t)\exp \left( i\,\mathbf{q}\cdot
\mathbf{r}\right) \ ,
\end{equation}
where, in the case of neutron scattering, cf.~Eq.~(\ref{a1-1}),
\begin{equation}
n(\mathbf{q},t)=\sum_{j}\exp \left[ -i\mathbf{q}\cdot \mathbf{R}%
_{j}(t)\right] \ .
\end{equation}
Since $n(\mathbf{r},t)$ is Hermitian, we have $n^{\dag }(\mathbf{q},t)=n(-%
\mathbf{q},t)$ and one obtains from Eq.~(\ref{a3})
\begin{equation}
W(t)=\lambda ^{2}\int_{0}^{t}dt^{\prime }\int_{0}^{t}dt^{\prime
\prime
}\sum_{f}\langle \psi _{f}|n(\mathbf{q},t^{\prime })\,\rho \,n(-\mathbf{q}%
,t^{\prime \prime })|\psi _{f}\rangle .  \label{a4}
\end{equation}

Assuming as usual  that $\Sigma _{f}|\psi _{f}\rangle \langle \psi
_{f}|=\mathbf{1}$ \cite{vHove,Squires} we get
\begin{equation}
\sum_{f}\langle \psi _{f}|n(\mathbf{q},t^{\prime })\,\rho \,n(-\mathbf{q}%
,t^{\prime \prime })|\psi _{f}\rangle =Tr\left[
n(\mathbf{q},t^{\prime })\,\rho \,n(-\mathbf{q},t^{\prime \prime
})\right] \ ,
  \label{a-ignore}
\end{equation}

If the integration in Eq.~(\ref{a4}) is extended over all times
(i.e., $t \rightarrow \infty$), this ensues over-all energy
conservation. This reproduces the well known result of standard
neutron scattering theory, cf.~\cite{vHove,Squires}.  Here,
however, it is important to retain the finite duration of the
scattering time, $t < \tau_{sc}$. This introduces an additional
freedom into the theory, because we may be able to observe the
influence of the decoherence on the scattering yield; see below.
The result will be expressed in terms of the correlation function
\begin{equation}
C(\mathbf{q},\tau )=Tr[n(\mathbf{q},t)\,\rho \,n(-\mathbf{q},t+\tau )]=Tr[n(%
\mathbf{q},0)\,\rho \,n(-\mathbf{q},\tau )] \  ,
  \label{a5}
\end{equation}
where we have utilized the fact that the scattering system is
stationary. By introducing the so-called {\it scattering time}
$\tau _{sc}$, representing the time interval in which  the
scattering process may happen, we find
\begin{equation}
W(\tau _{sc})  =  \lambda ^2\int\limits_0^{\tau _{sc}}dt^{\prime
}\int\limits_0^{\tau _{sc}}dt^{\prime \prime }C(q,t^{\prime \prime
}-t^{\prime })  =  \lambda ^2\int\limits_0^{\tau _{sc}}dt^{\prime
}\int\limits_0^{t^{\prime }}d\eta \left[ C(q,\eta )+C(q,-\eta
)\right] .
\end{equation}
Here we use the stationarity of the correlation function
\cite{Squires}. If we assume this function to be real, and that
$C(q,\eta )\approx 0$ for $\eta \gtrsim \tau _{sc},$ we obtain the
result
\begin{equation}
W(\tau _{sc})\approx 2\lambda ^2\tau _{sc}\int\limits_0^{\tau
_{sc}}d\eta \,C(q,\eta ).
\end{equation}
Now we can introduce the transition rate, $\dot{W}$ say, which is
defined as
\begin{equation}
\dot{W}\equiv \frac{W(\tau _{sc})}{\tau _{sc}}=2\lambda
^2\int\limits_0^{\tau _{sc}}d\eta \,C(q,\eta ).
 \label{a6}
\end{equation}
Here the correlation function is analogous to the so-called
intermediate function of neutron scattering theory \cite{Squires}.
This result for the scattering yield is analogous to that of
standard theory.

\subsection{Dynamics of open systems and scattering}

We now consider the scattering system to be open and strongly
interacting with its environment. We introduce a set of preferred
coordinates $\{\,|\xi \rangle \}$,
cf.~\cite{Petruccione,Mensky,Zeh}, representing the relevant
system's degrees of freedom coupled to the neutron probe. The
density matrix $\rho$ in (\ref{a5}) is then the \textit{reduced}
one in the space spanned by these states, and it is obtained by
tracing out the degrees of freedom of the "environment"
\cite{Petruccione,Mensky,Zeh}. For simplify, throughout this
section, let us denote this reduced density matrix by $\rho $ too.

In the subspace spanned by the preferred coordinates (also termed
"pointer basis"), let us assume a Lindblad-type equation
\cite{DECOH2}(a) of the form
\begin{equation}
\partial _t\rho =-i\left[ H,\rho \right] +\mathcal{R}\rho \equiv \mathcal{L}%
\rho  \label{b1}
\end{equation}
which has the formal solution  $\rho (t)=e^{\mathcal{L}t}\rho
(0)$.
   Let us look at a time-dependent expectation value
\begin{equation}
\langle A(t)\rangle \equiv Tr\left( \rho (t)A\right) =Tr\left( e^{\mathcal{L}%
t}\rho (0)A\right) =Tr\left( \rho (0)e^{\mathcal{L}^{\dagger
}t}A\right) \ ,
\end{equation}
where  $\mathcal{L}^{\dagger }$ is defined by
 $
Tr\left( \left( \mathcal{L}X\right) Y\right) =Tr\left( X\left( \mathcal{L}%
^{\dagger }Y\right) \right)
 $.
Thus it holds
$
\partial _{t}A(t)=\mathcal{L}^{\dagger }A(t)$;
cf.~\cite{DECOH2}(a). Here is  assumed that $\mathcal{L}$ is time
independent.

For the correlation functions like the one in Eq.~(\ref{a5}) it
then holds
\begin{equation}
\langle A(t)B\rangle =Tr\left[ \rho (0)\left(
e^{\mathcal{L}^{\dagger }t}A\right) B\right] =Tr\left[
Ae^{\mathcal{L}t}\left( B\rho (0)\right) \right] \equiv Tr\left(
A\rho _{B}(t)\right) ,
  \label{b2}
\end{equation}
where $\rho _{B}(t)$, as defined in Eqs.~(\ref{b2}), obeys the
equation
$
\partial _{t}\rho _{B}(t)=\mathcal{L}\rho _{B}(t)
$ with the initial condition
  $ \rho _{B}(0)=B\rho (0)$.
  Thus,
except for the initial condition, we have to solve the same
equation of motion as for the density matrix, Eq.~(\ref{b1}).

For simplicity of the further derivations, let us assume here a
simple Lindblad-type ansatz for the master equation having only
one Lindblad variable $X$. (In the real system we would have a
multitude of such variables.) Thus
\begin{equation}
\partial_{t}\rho =-i\left[ H,\rho \right] -K\left[ X,\left[ X,\rho \right]
\right] =\mathcal{L\rho } \ ,
  \label{c1}
\end{equation}
where the constant $K$ is real and $K>0$, $H$ is the reduced (or
relevant) Hamiltonian of a microscopic scattering system,  and the
double commutator term describes decoherence  (and/or dephasing).
For simplicity of the further calculations, we further assume that
we can take the preferred coordinates to commute with the total
Hamiltonian
\begin{equation}
H \mid\xi \rangle =  \mathcal{E}_{\xi } \mid\xi \rangle \ , \ \ \
\ \ \ \ \ \ X\mid\xi \rangle  =  \xi  \mid\xi \rangle   \ .
   \label{b3}
\end{equation}

This time evolution is now introduced into the correlation
function $C(\mathbf{q},\tau )$, Eq.~(\ref{a5}). A short
straightforward calculation (see Ref.~\cite{Aris+Stig} for full
details) yields
  for the transition rate the result
$$
 \dot{W}=  2
 \lambda ^{2}\int_{0}^{\tau _{sc}}\sum_{\xi ,\xi ^{\prime
}}\exp \left[ -i\left( \mathcal{E}_{\xi ^{\prime
}}-\mathcal{E}_{\xi }\right) \tau \right] \ \exp \left[ -K\left(
\xi ^{\prime }-\xi \right) ^{2}\tau \right] \times \ \ \ \ \
$$
\begin{equation}
\ \ \ \ \ \ \ \ \ \ \ \ \ \ \  \times \langle \xi \mid
n(-\mathbf{q},0)\mid \xi ^{\prime }\rangle \langle \xi ^{\prime
}\mid n(\mathbf{q},0)\,\rho (0)\mid \xi \rangle \, d\tau  \ .
\label{b-end}
\end{equation}

The decoherence-free limit of this result (i.e.  $K=0$)
corresponds to the conventional result of scattering theory. The
oscillatory terms $\exp \left[ -i\left( \mathcal{E} _{\xi ^{\prime
}}- \mathcal{E}_\xi \right) \tau \right] $ are due to the unitary
dynamics caused by the commutator part $-i\left[ H,\rho \right] $
of the master equation (\ref{c1}) for  $\rho $. These factors have
the absolute value 1.  If decoherence is present ($K>0$), and
especially if $K^{-1} \sim \tau _{sc}$,  the additional
contractive factors $\exp (-K\left( \xi ^{\prime }-\xi \right)
^2\tau )\leq 1$ can be seen to cause a decrease of the transition
rate and thus of the associated cross-section. This can be
illustrated  as follows.

Let us first assume that the reduced density operator $\rho (0)$
can be chosen to be \textit{diagonal} in the preferred $\xi
-$representation (which corresponds to the usual random phase
approximation). Then each term of Eq.~(\ref{b-end}) contains the
factor
\begin{equation}
\langle \xi |n(-\mathbf{q},0)|\xi ^{\prime }\rangle \langle \xi
^{\prime }|n (\mathbf{q},0)\,\rho (0)|\xi \rangle =|\langle \xi
\mid n(-\mathbf{q},0)\mid \xi ^{\prime }\rangle |^{2}\langle \xi
|\rho (0)|\xi \rangle \geq 0\ .
\end{equation}
 In the more general case with $\rho (0)$ being not diagonal in the
$\xi -$representation, one may proceed as follows. The decoherence
factors $\exp (-K\left( \xi ^{\prime }-\xi \right) ^{2}\tau )$
imply that only terms with $\xi \approx \xi ^{\prime }$ contribute
significantly to the transition
rate. Thus we may conclude that, by continuity, all associated terms with $%
\xi \approx \xi ^{\prime }$ in Eq.~(\ref{b-end}) should be
positive, too. The further terms with $\xi $ being much different
from $\xi ^{\prime }$ can be positive or negative. But they may be
approximately neglected, since they decay very fast and thus
contribute less significantly to $\dot{W}$; cf.~\cite{Mensky}.

The main conclusion from the preceding considerations is that the
time average over $\tau _{sc}$ in Eq.~(\ref{b-end}) always
\textit{decreases}  the value of
 $\dot{W}
\equiv W(\tau _{sc})/\tau _{sc}$, due to the presence of the
contractive factors $\exp (-K\left( \xi ^{\prime }-\xi \right)
^2\tau )\leq 1$. In other words, the effect of decoherence (and/or
dephasing) during $\tau _{sc}$ plays a crucial role  and may lead
to an "anomalous" decrease of the transition rate and the
associated scattering intensity. This result is in line with that
of Ref.~\cite{LaserPhysics}, which investigated the standard
expression of the double differential cross-section of neutron
Compton scattering theory \cite{Sears,Watson} by assuming
decoherence of final and initial states of the scattering system.

For further illustration let us consider the following two
specific limiting cases: $(A)$ For "vanishing" decoherence,
$K\rightarrow 0$, the above contractive factors go to 1 and thus
the anomalous scattering effect disappears. I.e. the preceding
result (\ref{b-end}) agrees with the conventional theoretical
results \cite{Watson,Squires}. $(B)$ In the opposite  case,
$K\rightarrow \infty$, only the "diagonal" terms with $\xi=\xi
^{\prime }$ survive in Eq.~(\ref{b-end}) and the related
contractive factors go to 1;  additionally, the oscillating
factors become equal to 1. Consequently, the time-integration in
Eq.~(\ref{b-end}) has no effect and the rhs of this equation goes
over to the standard expression Eq.~(\ref{a6}). Also this result
is in line with conventional expectations.

\section{Conclusions}

Theoretical considerations suggest the presence of attosecond
quantum entanglement of the scattering protons and the surrounding
electrons \cite{PRL97,PRL03,EPL99,JACS,LaserPhysics}. Furthermore
the usual Born-Oppenheimer approximation is not applicable
\cite{JACS,Nikitas,LaserPhysics,Reiter}. Moreover, recent NCS
results from liquid HD and the equimolar H$_2$-D$_2$ mixture
showing identical anomalous scattering, provided strong evidence
that quantum exchange correlations between proton pairs cannot be
the main physical reason for the effect \cite{HD}. For further
theoretical discussions, see \cite{LaserPhysics}.

In contrast to standard scattering theory  \cite{vHove,Squires} of
thermal and/or cold neutrons, in which  decoherence plays no role
at all, our theoretical treatment of NCS  is based on the physical
fact that micro- and/or mesoscopic scattering systems are open
quantum systems. This was shown to follow from the ultrashort
scattering time of NCS and ECS. The revealed "anomalous" effect,
which has no conventional interpretation, indicates that
attosecond entanglement (and its decoherence) involving protons
are quantum phenomena of broader significance and relevance  than
realized so far.

 \section*{Acknowledgments}
 This work was supported, in parts, by the
 EU (network QUACS)  and by a grant from the
Royal Swedish Academy of Sciences.

 The author would like to thank T. Abdul-Redah, M. Arndt, E.
Br\"andas, D. Colognesi, F. Fillaux, N. Gidopoulos, B. Hessmo, E.
Joos, E. B. Karlsson, C. Kiefer, G. Kurizki, J. Manz, J. Mayers,
I. E. Mazets, M. Mensky, H. Sillescu, S. Stenholm and M. Vos for
various helpful and insightful discussions, and ISIS for generous
beam time allocation.

\vfill\eject
\end{document}